\documentclass{article}
\usepackage{graphics}
\begin{document}
\title{The Theory of SNOM: A novel approach}
\author{A. J. Ward, J. B. Pendry\\ The Blackett Laboratory, Imperial College,\\
London, SW7 2BZ, UK}
\maketitle

\begin{abstract}
In this paper we consider the application of electromagnetic theory to
the analysis of the Scanning Near-field Optical Microscope (SNOM) in
order to predict experimentally observable quantities such as the
transmission or reflection coefficients for a particular tip-surface
configuration. In particular we present the first application of a
transfer matrix based calculation to this challenging problem by using
an adaptive co-ordinate transformation to accurately model the shape of
the SNOM tip.

We also investigate the possibility of increasing the transmitted light
through the SNOM tip by introducing a metal wire into the centre of the
tip. This converts the tip into a co-axial cable. We show that, in
principle, this can dramatically improve the transmission characteristics
without having a detrimental effect on the resolution.

\end{abstract}
\section{Introduction}

One of the first lessons in optics which is drummed into our
heads is that the resolution of any
optical instrument is limited, fundamentally, by the wavelength of the
light it uses. 
However, this resolution limit is not an absolute and, in fact, applies only
to the 
image in the far field. Sub-wavelength information can, in
principle, be obtained by probing the near field which contains both the
propagating modes  and evanescent
modes bound to the surface of the object. It is these evanescent modes
which correspond to sub-wavelength spatial frequencies and so encode the
nano-metric detail of the object. 
So the principle which lies behind near-field imaging techniques is this: find
some way to probe the near-field and imaging details smaller than
the wavelength becomes feasible. 

In the early nineties, Pohl and co-workers at
IBM~\cite{Pohl,Pohl2}, designed a near-field `optical stethoscope' for imaging
at a wavelength 
of 448nm with a claimed resolution of about $\lambda/20$. 
Since then, many other groups have built Scanning Near-field Optical
Microscopes (SNOM, or sometimes NSOM) along similar lines. A typical, modern 
SNOM can
be described as follows. The surface under investigation is illuminated from
above by laser light passing down an optical fibre. The end of the fibre is
formed into a taper and the sides of the taper are coated in
about  100nm of a 
metal, typically aluminium or silver~\cite{Betzig}, the very end of the tip
being left uncoated. In this way 
apertures can be reliably produced with diameters of $\sim$ 20-500nm. As the
tip is scanned over the surface, a 
shear-force mechanism is used to regulate the tip surface
distance~\cite{Crow,Betzig3}. At each point the light transmitted through the
sample is collected by a 
photomultiplier and a complete image built up.
Such a device also offers the possibility for performing local, high 
resolution spectroscopy~\cite{Trautman,Xie,Ambrose}.

Many variations on this basic idea are possible and there are several
comprehensive reviews~\cite{snomrev,Courjonrev}. Possibilities 
include measuring the reflected
rather than the transmitted light, illuminating from below and using the coated
fibre tip as a light collector, or even illuminating from below above the
critical angle and using an uncoated tip as a nano-scatterer to disrupt the
near-field and frustrate total internal
reflection~\cite{Courjon,Reddick}. 

However, if the SNOM is to mature as an analytical tool then theorists have
a critical role to play, both in order to interpret correctly the images
produced by the microscope, but also to assist in the optimisation of
tip and illumination system. Several approaches are possible for solving
the problem of the electromagnetic scattering between
tip and surface, amongst those suggested have been finite difference
time domain schemes and boundary matching methods. The former usually
require a super-computer to handle a problem as complex as this. The
latter, which expands the fields in each region in some suitable
complete basis
set and matches the fields at the boundaries, faces difficulties when
handling regions of irregular shapes, which are needed for a realistic
description of the tip. 
However, one group have completely dominated the theory of the
near-field microscope by introducing an iterative Green's function based
method~\cite{Girard2,Girard3}. 
Using this method Girard {\it etal.}\ have been able to simulate images
produced by reflection microscopes scanning over a range of
dielectric and metal objects.

\section{A novel approach}

However, despite the sophistication of Girard's method it does rely on a
real space discretisation in order to represent both object
and tip. For an accurate description of the tip this is can be a serious
restriction, especially if we are restricted to a regular, Cartesian grid. The
problem is this: in reality the surface of the tip is a cone or paraboloid
shape with a smooth surface, discretisation leads to a surface covered
in steps, the size of which depends on the mesh size. The
presence of these steps will result in additional scattering from the
surface of the tip and may reduce the accuracy of the calculation
especially if the size of the discretisation mesh is large. The only way
to reduce the error is to use more mesh points leading to larger, slower
calculations. 

We can, however, tackle the SNOM problem from a slightly different angle which
helps us to overcome these problems. The idea is to perform the discretization
in a co-ordinate system which follows the shape of the tip itself. While, in
principle, the construction of such an adaptive, discrete mesh may seem a
horribly complex task, in practise this is not so. This is because of a useful
result which allows us to factorise the problem into two parts. This result,
which is derived in detail elsewhere~\cite{nonorth}, allows us to transfer all
the 
details of the mesh geometry into an effective, tensorial permittivity and
permeability ($\hat{\varepsilon}^{\alpha\beta}$,
$\hat{\mu}^{\alpha\beta}$). Now we need only perform the discretization on a
regular, cartesian mesh and then map that onto the adaptive mesh by
introducing the appropriate $\hat{\varepsilon}$ and $\hat{\mu}$.

How do Maxwell's equations appear in the adaptive co-ordinate system? The
result is surprisingly simple. Let $q_{1}, q_{2}, q_{3}$ be the co-ordinates
in our adaptive co-ordinate system and let the unit vectors $\mathbf{u}_{1},
\mathbf{u}_{2}, \mathbf{u}_{3}$ point along the generalised $q_{1}, q_{2},
q_{3}$ axes. In this system Maxwell's equations become:
\begin{equation}
\nabla _q\times \hat {\bf E}=-\mu _0\hat \mu \;\partial \hat
{\bf H}/\partial t
\end{equation}
\[
\nabla _q\times \hat {\bf H}=+\varepsilon _0\hat \varepsilon \;\partial
\hat {\bf E}/\partial t
\]
where,
\[
\hat{\varepsilon}^{\alpha\beta}=\varepsilon\;g^{\alpha\beta}\left| {\bf
u_1\cdot 
\left( {\bf u_2\times
\bf u_3} \right)} \right|\;\frac{Q_1Q_2Q_3}{Q_{\alpha}Q_{\beta}}
\]
\begin{equation}
\hat \mu ^{\alpha\beta}=\mu \;g^{\alpha\beta}\left| {\bf u_1\cdot \left( {\bf
u_2\times \bf u_3} \right)} \right|\;\frac{Q_1Q_2Q_3}{Q_{\alpha}Q_{\beta}}
\end{equation}
\[
\hat{E}_{\alpha}=Q_{\alpha}E_{\alpha}\hspace{1cm}
\hat{H}_{\alpha}=Q_{\alpha}H_{\alpha}
\]
and,
\begin{equation}
Q_{\alpha}^{2}=\left(\frac{\partial x}{\partial q_{\alpha}}\right)^{2}
+\left(\frac{\partial y}{\partial q_{\alpha}}\right)^{2}
+\left(\frac{\partial z}{\partial q_{\alpha}}\right)^{2}
\end{equation}
\begin{equation}
{\bf g}^{-1}
=\left[ {\begin{array}{ccc} {\bf u_1\cdot \bf u_1}&{\bf u_1\cdot \bf u_2}&{\bf
u_1\cdot \bf u_3}\\ {\bf u_2\cdot \bf u_1}&{\bf u_2\cdot \bf u_2}&{\bf
u_2\cdot \bf u_3}\\ {\bf u_3\cdot \bf u_1}&{\bf u_3\cdot \bf u_2}&{\bf
u_3\cdot \bf u_3}
\end{array}}\right]
\end{equation}
Notice that in general $\mathbf{u}_{\alpha}$, $Q_{\alpha}$ and
 $g^{\alpha\beta}$ as well as $\varepsilon$ and $\mu$ will be functions of
position. 

\section{Details of our calculations}

We wish to calculate the transmitted intensity through a given tip and surface
configuration and to this end we shall employ a transfer matrix method familiar
to the theory of photonic band gap materials~\cite{J+A}. Put simply this
involves 
integrating the electric and magnetic fields from one side of the system to
the other by means of a \textit{transfer matrix} which relates the fields in
one layer of the mesh to the fields in the next layer.

\begin{figure}[tb]
\centering
 \hrule
 \vspace{1cm}
 \resizebox{\textwidth}{!}
 {\includegraphics{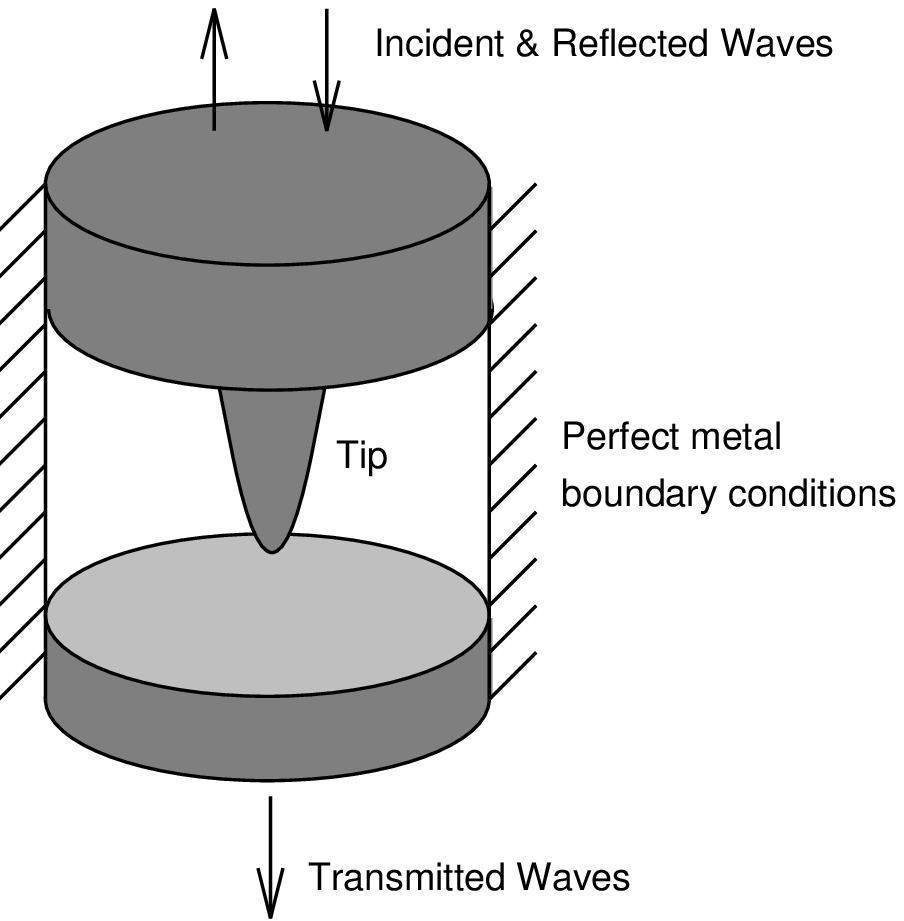}}
 \vspace{1cm}
 \hrule
 \caption{The tip-surface scattering system is contained within a
cylinder. Waves can be incident from top or bottom, but the boundary
conditions at the side must be specified.}
 \label{fig:tipBC}
\end{figure}

As a first attempt calculation we will try to simplify the problem as
much as possible. By exploiting the obvious cylindrical symmetry of the
tip we can reduce the problem from a three dimensional to a two
dimensional one. The other major simplification we can impose is in the
boundary conditions. By imposing cylindrical symmetry we have confined
the calculation to a cylindrical region of space containing both tip and
surface. As usual for a transfer matrix method we can send an incident
wave into the system from either end of the cylinder and find the
transmission and reflection coefficients~\cite{JBP} but we must specify the
boundary conditions for the fields at the cylinder's edge
(see figure~\ref{fig:tipBC}). The simplest form these boundary conditions
can take is if we enclose the cylinder with a perfect metal, in other
words we set the fields to zero at the boundary. This is not a trivial
approximation to make, a much better solution would be to expand the
fields outside the scattering region in some complete basis set and
match the fields at the boundary, but it is not a foolish approximation
either. 
As long as the wavefield does not have a sizable magnitude at the boundary and 
remains tightly localised between the tip and sample, as we might expect it to,
then the choice of boundary conditions should not have a major effect on the
results.

\subsection{An analytic form for the tip}

Our next choice is that of the discretisation mesh itself for this will
determine the shape of the tip which we model. Ideally, we wish to
specify the mesh geometry by means of an analytic function -- this will
greatly simplify the process of defining the effective permittivity and
permeability, $\hat{\varepsilon}$ and $\hat{\mu}$, which will contain
all the information about the co-ordinate system. We also need to
include sufficient parameters in the chosen function to allow us to fine
tune the shape of the tip.

\begin{figure}[tb]
 \hrule
\resizebox{\textwidth}{!}
{\includegraphics{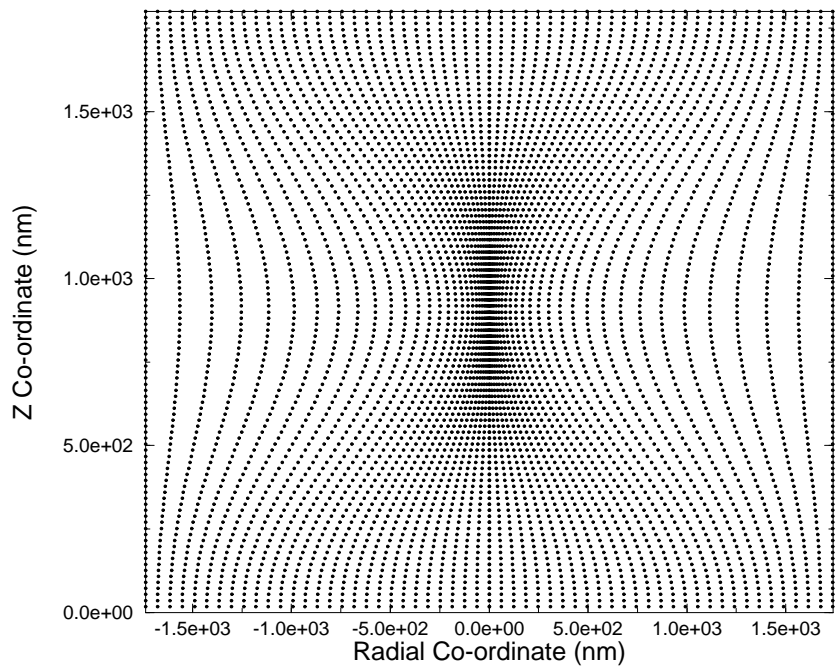}}
 \hrule
 \caption{A typical mesh used to represent the SNOM tip.}
 \label{fig:tipmesh}
\end{figure}

One possible form for the mesh is this:
\begin{eqnarray}
r&\hspace{-3mm}=&\hspace{-3mm} a \,I\!R + \frac{1}{2}\left[
1-\cos\left(\frac{2\pi(I\!Z-1)}{I\!Z_{max} -1}\right)\right] \{
\delta-a\} I\!R\left(1-\frac{I\!R}{I\!R_{max}}\right)^{\alpha} \\
\theta&\hspace{-3mm}=&\hspace{-3mm}b \,I\!\Theta \\
z&\hspace{-3mm}=&\hspace{-3mm}c \,I\!Z
\end{eqnarray}
where $I\!R$, $I\!\Theta$ and $I\!Z$ are the discrete variables which
label each point on the mesh. This form contains five free parameters.
The first three $a,b,c$ describe an undistorted mesh. That is to say if
$\delta=a$  and $\alpha=1$ the mesh defined is just discrete cylindrical
polar co-ordinates with a radial spacing of $a$, angular spacing of $b$
and a spacing $c$ in the z-direction. The other two parameters
$\alpha$ and $\delta$ control the distortion of this cylindrical mesh
into a kind of hour glass shape; $\delta$ defines the minimum radial
distance between points at the centre of the mesh and $\alpha$ controls
how tightly the other co-ordinate lines bunch in at the centre.
Figure~\ref{fig:tipmesh} shows the shape of the co-ordinate lines for a
typical choice of parameters.

\begin{figure}[tb]
 \hrule
\resizebox{\textwidth}{!}
{\includegraphics{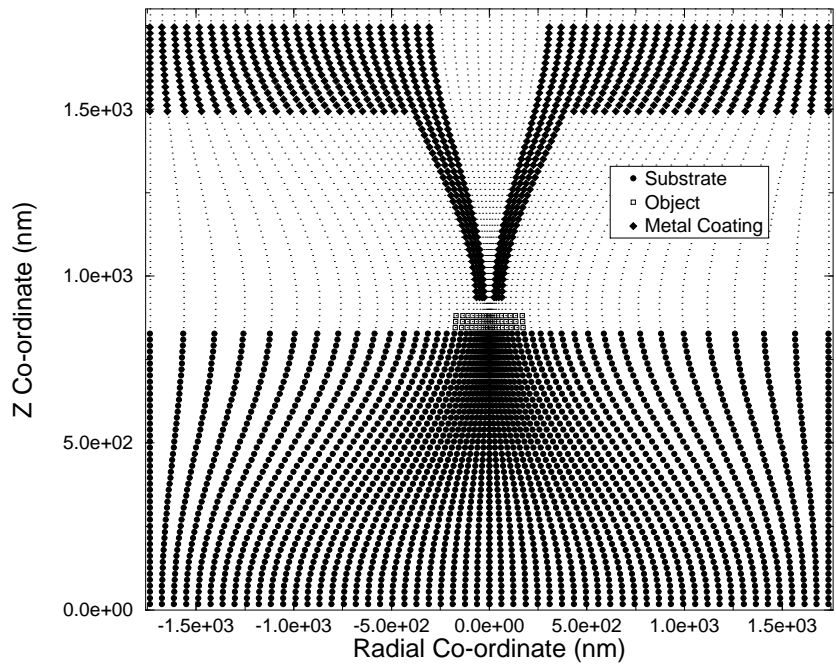}}
 \hrule
 \caption{The mesh shaded to indicate the presence of the metal coating the 
tip (Diamonds), the substrate (circles) and the object being studied (squares)}
 \label{fig:tipmesh_fill}
\end{figure}

The tip itself can now be defined by assigning the dielectric constant
at each point in the mesh. The tip and substrate are glass,
$\varepsilon=(2.16,0.0)$ the tip is coated with a 90nm layer of aluminium
$\varepsilon=1-\omega_{p}^{2}/ \omega(\omega+i\gamma)$ with
$\omega_{p}=15.1$eV and $\gamma=0.27$eV.
The very end of the tip is not coated and this forms the aperture.
A layer of metal around the top of the tip prevents stray light
from the source leaking around the side of the tip. 
Any object we wish to image can be introduced by changing the dielectric
constant just above the surface. Figure~\ref{fig:tipmesh_fill} shows how
the metal coating layer, substrate and object fit onto the distorted 
co-ordinate system.

\section{Results}

\begin{figure}[tb]
\centering
 \hrule
 \vspace{1cm}
 \resizebox{\textwidth}{!}
 {\includegraphics{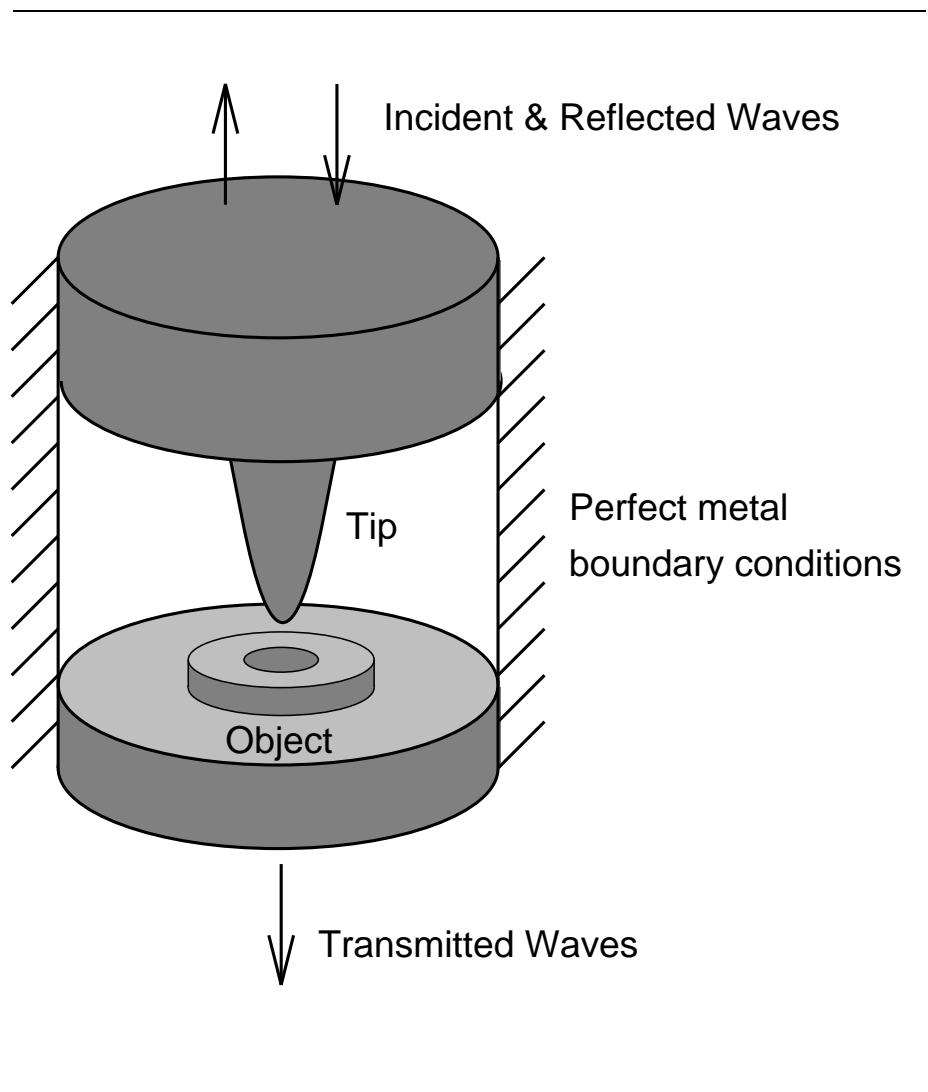}}
 \vspace{1cm}
 \hrule
 \caption{Disc or ring objects (ring shown) can be placed under the tip.}
 \label{fig:tip_obj}
\end{figure}

We will want to make comparisons between different tip designs to see if
we can suggest ways to improve the utility of the near-field microscope.
In order to do this we shall need some way to estimate the resolution of
a given microscope configuration as well as to calculate the transmitted
or reflected signal. One way to test the resolution is to
compare the signals for two similar objects. 
Ideally, we would like to calculate the transmission for the tip in
arbitrary positions above an arbitrary object and in this way build up a
complete image of the object. Unfortunately, we have constrained
ourselves somewhat by insisting upon cylindrical symmetry so this is not
possible. The only choices available to us for the objects under the tip
are rings and discs with the tip sitting directly above their centre.
This is made clear in figure~\ref{fig:tip_obj}.
Fortunately this is
enough to get an estimate of the resolution, as we can compare a ring
and a disc of the same outer radius to see whether we can detect the
presence of the hole at the centre of the ring.

\begin{figure}[tb]
\hrule
\resizebox{\textwidth}{!}
{\includegraphics{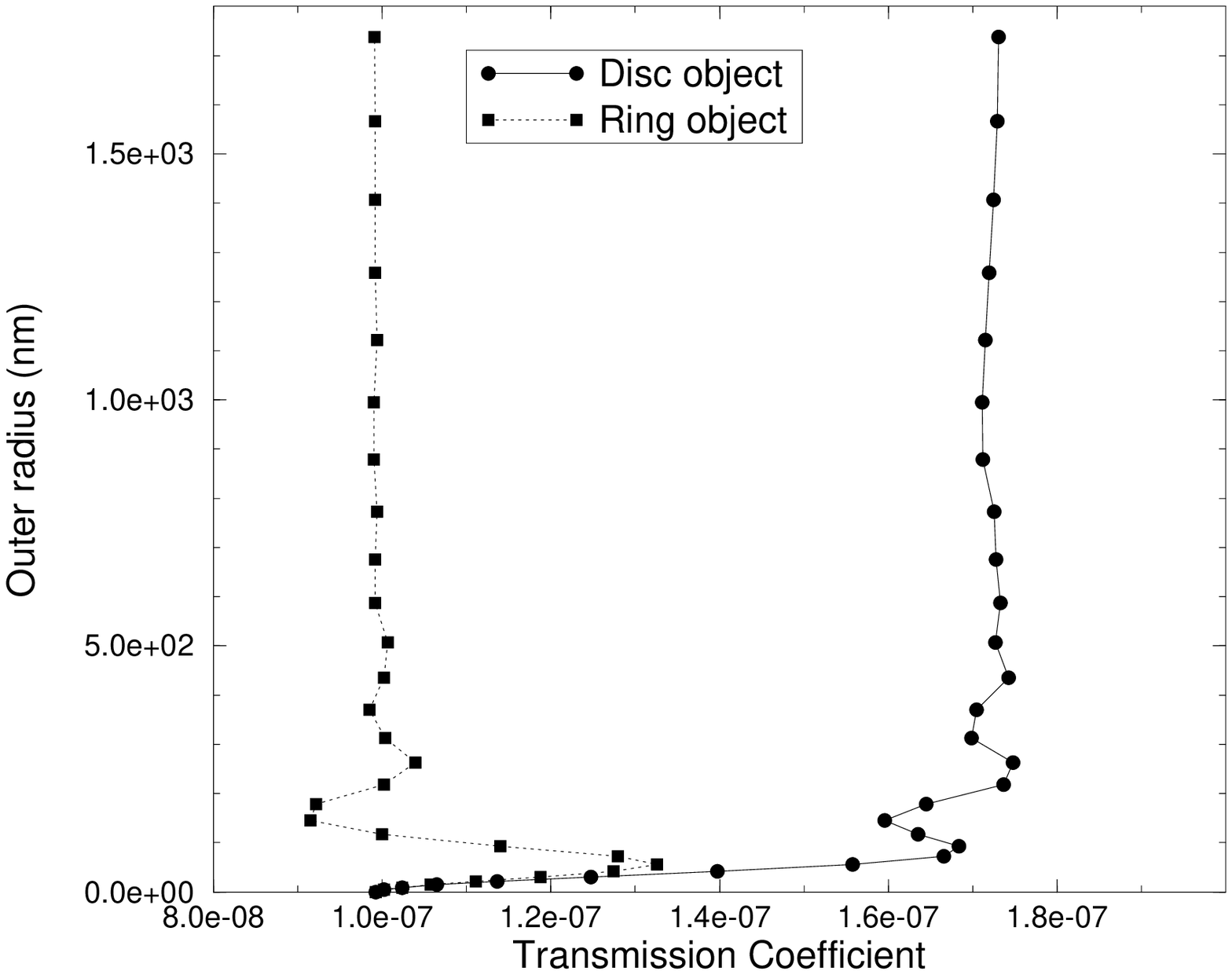}}
\hrule
 \caption{Comparison of transmission coefficients for ring and disc
objects. Aperture radius  48nm}
  \label{fig:tip1}
\end{figure}

\begin{figure}[tb]
\hrule
\resizebox{\textwidth}{!}
{\includegraphics{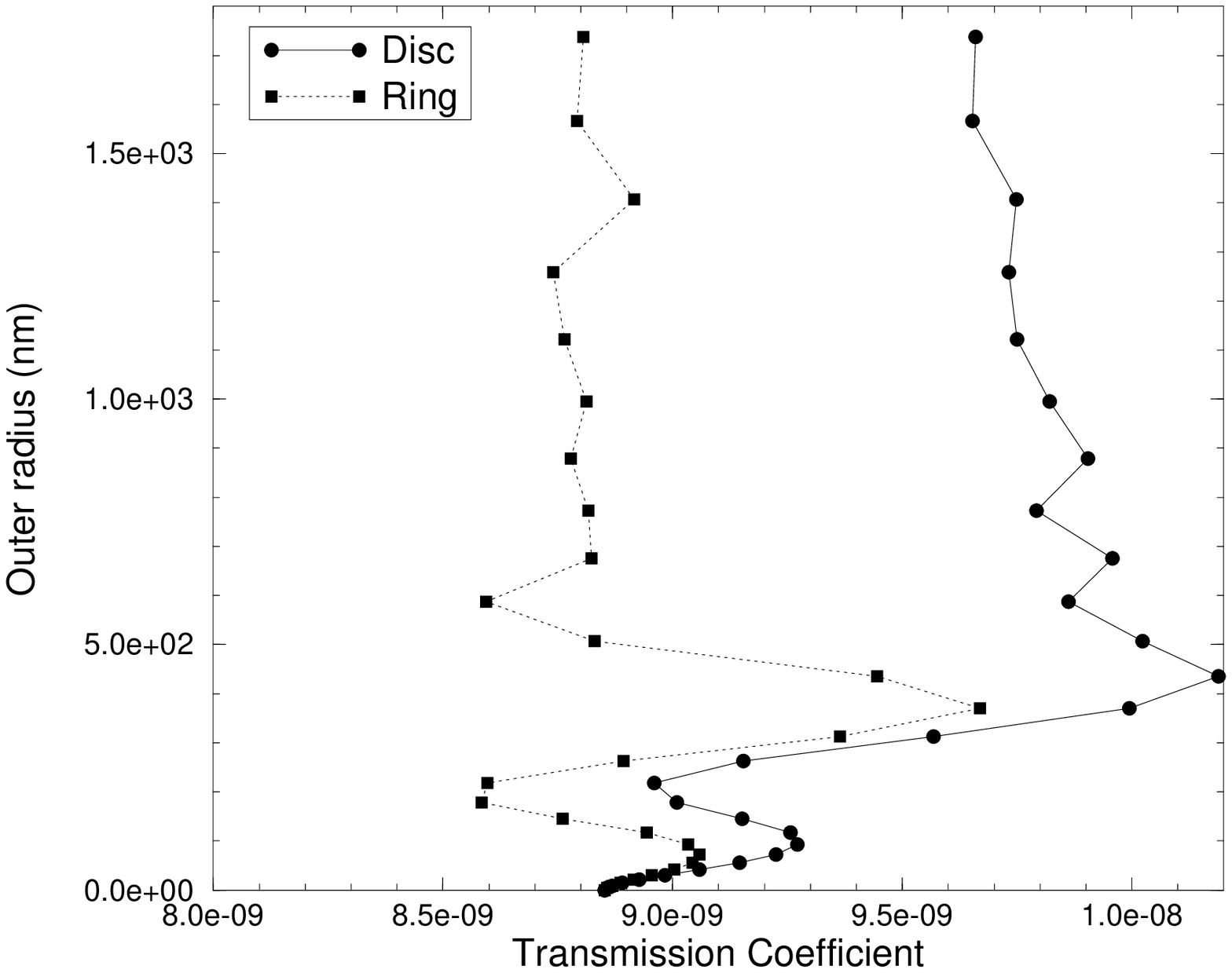}}
\hrule
 \caption{Comparison of transmission coefficients for ring and disc
objects. Aperture radius 35nm}  \label{fig:tip2}
\end{figure}

\begin{figure}[tb]
\hrule
\resizebox{\textwidth}{!}
{\includegraphics{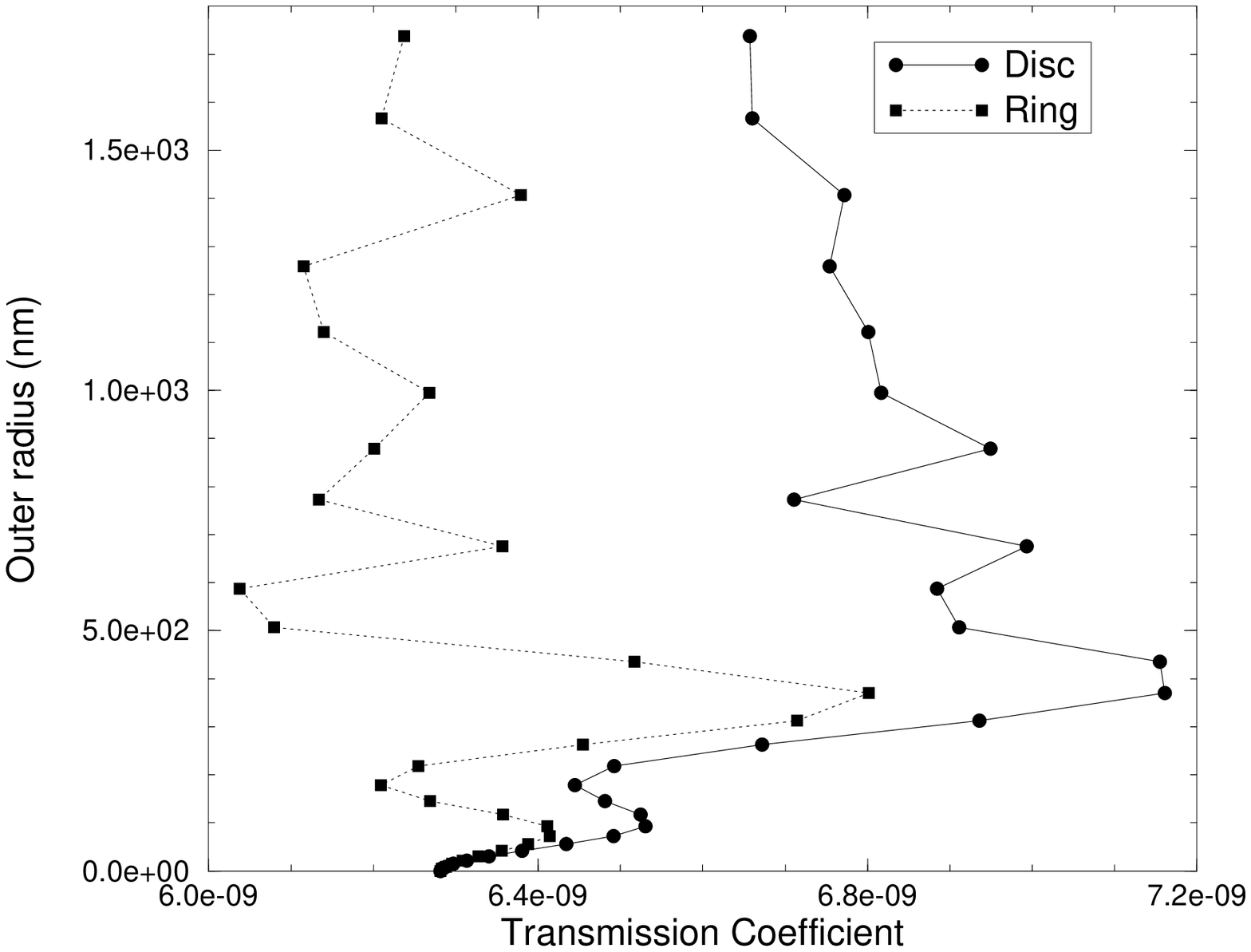}}
\hrule
 \caption{Comparison of transmission coefficients for ring and disc
objects. Aperture radius 25nm}  \label{fig:tip3}
\end{figure}

Figures~\ref{fig:tip1} to~\ref{fig:tip3} show comparisons between
transmission coefficients for
ring and disc objects for three different tip apertures, given a fixed
incident wave of frequency 2.5eV.
The objects are made of the same glass as the substrate, 54nm high and
the tip-object separation is fixed at 36nm.
Each graph gives the transmission against the outer radius of the
object. The inner radius of the ring
is a few nanometres smaller than the outer radius. Strictly, the
outer radius is one radial mesh point larger than the inner radius so,
because the mesh points are more closely packed at the centre, 
the difference increases as the size of the object increases. 

As the aperture decreases the overall
magnitude of the transmission coefficient is reduced, from around $10^{-7}$
for an aperture radius of 48nm down to about $6\times10^{-9}$ for a 25nm
aperture. Reducing
the aperture further proved difficult as the magnitude of the transmission
coefficient 
is reduced so much that the variations in it become comparable with the
rounding  
errors in the calculation.

\subsection{Improving the design: Coaxial tips}

In order to get a high resolution, clearly what is required is as small
an aperture as possible. However, because all the guided modes of
the tip run into cut-off once the diameter is smaller than the
wavelength, small apertures come at the price
of a very small transmitted signal. 
This is borne out by the results in
the previous section, where the transmission for an aperture of 25nm is
about two orders of magnitude down on the transmission for a 48nm aperture.

However there is, in principle, a way to overcome this problem.
Inserting a wire into the centre of the tip would convert the metal clad
waveguide into a tapered coaxial cable. The coax is well known to have
a 
different kind of guided mode called the TEM mode~\cite{jackson}
arising from the different boundary conditions satisfied at the centre.
This mode has a free space like dispersion with no cut-off and so will
propagate down the cable for any value
of its radius offering the possibility of delivering a much higher photon
flux to the end of the tip.

There are one or two potential problems with using a coaxial tip in a
microscope, 
of course. The main
one is the fear that the extra absorption caused by introducing a lossy
metal into the centre of the guide would negate any gain given by the
new mode, especially as the field distribution in the TEM mode behaves like
$1/r$ 
so is largest along the central wire.
We do our best to minimise this by limiting the central wire
only to the very end of the tip, where the radius has fallen below the
cut-off point. We also try to maintain a uniform impedance along the
coaxial part by ensuring that the central wire tapers in the same
proportions as the rest of the tip. The other constraint we must be
careful with is not to make the wire too thin. Once the radius of the
wire is much less than the skin depth it will no longer be effective.

\begin{figure}[tb]
\hrule
\resizebox{\textwidth}{!}
{\includegraphics{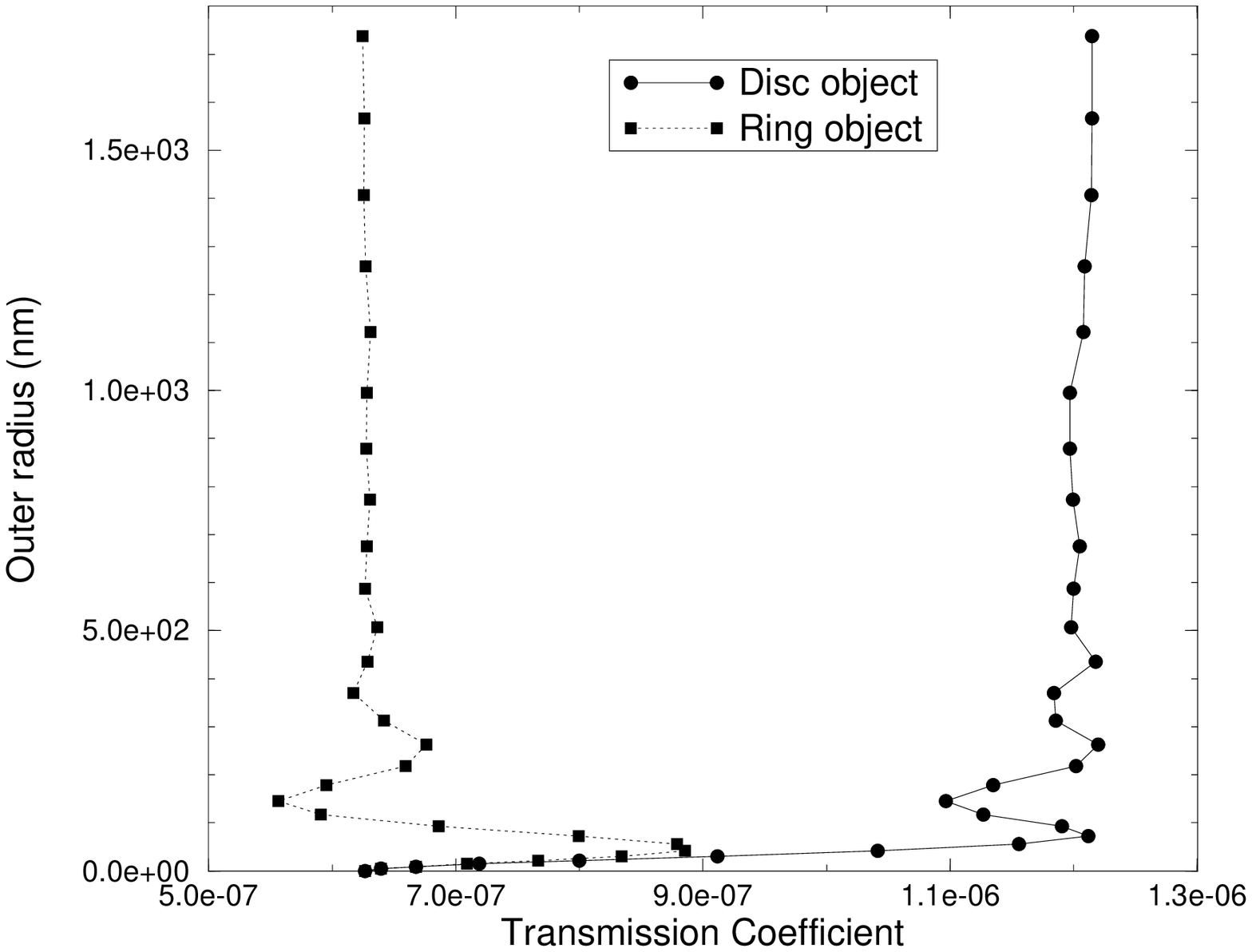}}
\hrule
 \caption{Comparison of transmission coefficients for ring and disc
objects. New Coaxial tip of aperture 35nm}  \label{fig:coax1}
\end{figure}

\begin{figure}[tb]
\hrule
\resizebox{\textwidth}{!}
{\includegraphics{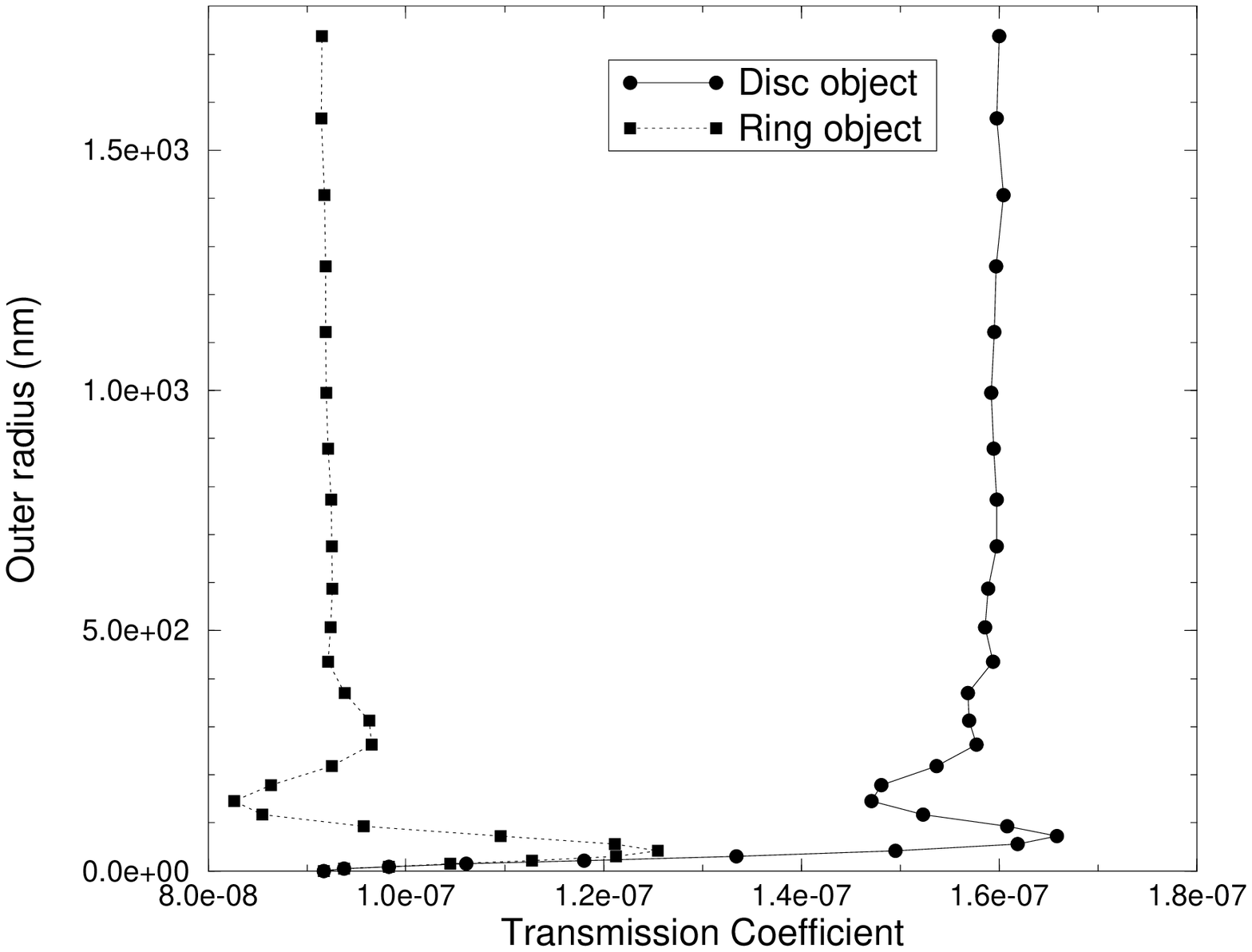}}
\hrule
 \caption{Comparison of transmission coefficients for ring and disc
objects. New Coaxial tip aperture 25nm}  \label{fig:coax2}
\end{figure}

\begin{figure}[tb]
\hrule
\resizebox{\textwidth}{!}
{\includegraphics{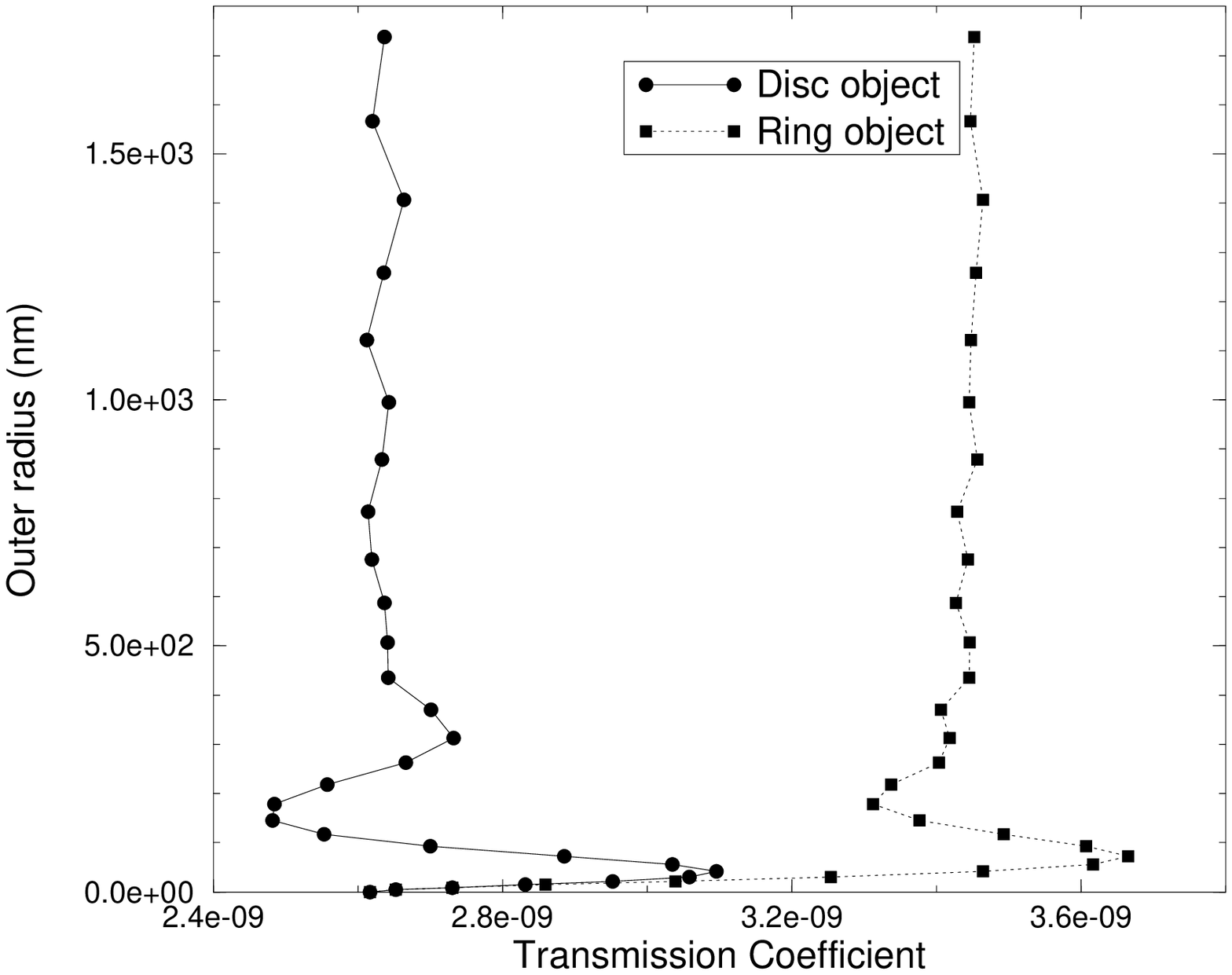}}
\hrule
 \caption{Comparison of transmission coefficients for ring and disc
objects. New Coaxial tip aperture 17nm}  \label{fig:coax3}
\end{figure}

In order to optimise the transmitted power through the coaxial tip we sought a
compromise between the lack of cut-off in the TEM mode and the increased
absorption 
due to the central wire. To this end we varied the length of wire inside
the end 
of the tip and found the optimum value to be 180nm. The radius of the wire at
the  
very end of the tip is 6.8nm.  
We then repeated the previous calculations, the results being
shown in figures~\ref{fig:coax1} to~\ref{fig:coax3}.

Comparing figures~\ref{fig:coax1}
to~\ref{fig:coax3} to figures~\ref{fig:tip1} to~\ref{fig:tip3} shows
the improvement in the magnitude of the transmission when
using the new tip.
For the 35nm aperture the improvement is about two orders of magnitude, falling
to about one order of magnitude for the 25nm aperture.
The general increase in signal strength has allowed us to perform the
calculation 
for a smaller aperture than before, of only 17nm radius.
Also it is clear that the resolution for the coaxial tip is equal to if not
slightly 
better than the conventional tip. 
The shape of the graphs at larger object radii is clearly much flatter
than for the normal tip. We expect the fields to be strongly confined in
the region between tip and sample. The more tightly the fields are
confined, the smaller the spot on the surface illuminated and so the
less effect adding dielectric material to the object at large radii has
on the transmission. This would seem to suggest fields more tightly
confined for the coaxial tip compared to the normal one.
 
\begin{figure}[tb]
\hrule
\resizebox{\textwidth}{!}
{\includegraphics{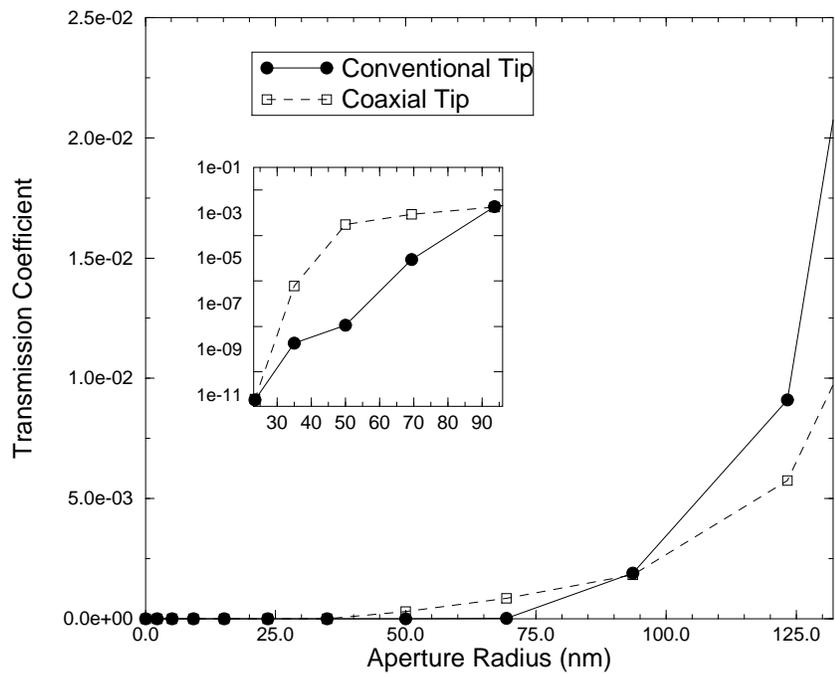}}
\hrule
 \caption{Transmission coefficient against aperture radius for the
conventional and coaxial tips. The inset is a plot of the log of the
transmission 
against radius in the critical region}  \label{fig:enhance}
\end{figure}

Figure~\ref{fig:enhance} shows the transmission coefficient as a
function of 
aperture radius for both types of tip and shows that for the range of
apertures 
from about 65nm to 20nm the coaxial design offers significant improvement.

\section{Conclusions and future work}
In this paper we have presented the first attempt to apply the transfer
matrix 
method to calculate the properties of the near-field optical microscope. We
have  
calculated the transmission coefficient for a range of aperture radii as well
as  
estimating the resolution by comparing the transmission data for slightly
different 
objects under the tip. 

Whilst our calculations may lack the sophistication of others that have been
 performed
in this field, ours are the first, to the best of our knowledge, to use an
adaptive 
mesh scheme to follow accurately the shape of the tip. There is no reason, in 
principle, why this co-ordinate transformation cannot be applied to other
methods. 
Many  other techniques such as Girard's method or the Finite difference
time domain method rely on a
real space discretisation at some point and so could directly benefit from the 
adaptive mesh we have developed here.
In the future we also hope to remove the constraint of working within 
cylindrical symmetry and so calculate images of objects directly.
 
We have also presented the first calculations for a near-field microscope with
a 
coaxial tip. These results clearly show the potential for such a tip to
enhance the 
magnitude of the transmitted signal even at very small aperture sizes. The
only  
theoretical limit on the size of such a tip is the need for the central wire
to have 
a diameter larger than the skin depth.

Experimentally, some progress has been made towards realising such a tip.
Fischer {\it etal.}~\cite{coax} investigated the properties of a conical
glass tip manufactured with a silver wire taper placed at the
centre. However, difficulties 
in obtaining an unbroken metal core prevented him from using the tip for
imaging. More recently, he has suggested a tetrahedral tip~\cite{Fischer2}
coated with 
metal except for one edge as an alternative which may offer the advantages of
the coaxial design but without the manufacturing difficulties. It is hoped
that calculations of the kind outlined here can be of use in analysing the
merits of alternative SNOM designs such as these.




\begin{thebibliography}{99}
\bibliographystyle{plain}
 \newcommand{\apl}[1]{{\it Appl.\ Phys.\ Lett.} {\bf #1}}
 \newcommand{\ao}[1]{{\it Applied\ Optics} {\bf #1}}
 \newcommand{\prb}[1]{{\it Phys.\ Rev.\ B} {\bf #1}}
 \newcommand{\prl}[1]{{\it Phys.\ Rev.\ Lett.} {\bf #1}}
 \newcommand{\prv}[1]{{\it Phys.\ Rev.} {\bf #1}}
 \newcommand{\jpcm}[1]{{\it J.\ Phys.:\ Condens.\ Matter} {\bf #1}}
 \newcommand{\jpc}[1]{{\it J.\ Phys.\ C} {\bf #1}}
 \newcommand{\ssc}[1]{{\it Solid\ State\ Commun.} {\bf #1}}
 \newcommand{\epl}[1]{{\it Europhys.\ Lett.} {\bf #1}}
 \newcommand{\zpb}[1]{{\it Z.\ Phys.\ B} {\bf #1}}
 \newcommand{\jmo}[1]{{\it J.\ Mod.\ Optics } {\bf #1}} 
 \newcommand{\josab}[1]{{\it J.\ Opt.\ Soc.\ Am.\ B} {\bf #1}} 
 \newcommand{\ol}[1]{{\it Optics Lett.} {\bf #1}}
 \newcommand{\jap}[1]{{\it J.\ Appl.\ Phys.} {\bf #1}}
 \newcommand{\nat}[1]{{\it Nature} {\bf #1}}
 \newcommand{\sci}[1]{{\it Science} {\bf #1}}
 \newcommand{\oc}[1]{{\it Optics\ Commun.} {\bf #1}}

\bibitem{Pohl} Pohl D W, Denk W, Lanz M, 1984, \apl{44} 651.
\bibitem{Pohl2} D\"{u}rig U, Pohl D W, Rohner F, 1986, \jap{59} 3318.
\bibitem{Betzig} Betzig E, Trautman J K, Harris T D, Weiner J S, Kostelak R L,
1991, \sci{251} 1468.
\bibitem{Crow} Toledo-Crow R, Yang P C, Chen Y, Vaez-Iravani M, 1992,
\apl{60} 2957.
\bibitem{Betzig3} Betzig E, Finn P L, Weiner J S, 1992, \apl{60} 2484. 
\bibitem{Trautman} Trautman J K, Macklin J J, Brus L E, Betzig E, 1994,
\nat{369} 40.
\bibitem{Xie} Xie S X, Dunn R C, 1994, \sci{265} 361.
\bibitem{Ambrose} Ambrose W P, Goodwin P M, Martin J C, Keller R A,
1994, \sci{265} 364.
\bibitem{snomrev} Courjon D, Baida F, Bainier C, Van Labeke D, Barchiesi
D, 1995, `Photons and Local Probes' 59-77,  Ed. Marti O, M\"{o}ller R, Kluwer
Academic Publishers. 
\bibitem{Courjonrev} Courjon D, Bainier C, 1994, {\it
Rep.\/ Prog.\/ Phys.\/ } {\bf 57} 989.
\bibitem{Courjon} Courjon D, Sarayeddine K, Spajer M, 1989, \oc{71} 23.
\bibitem{Reddick} Reddick R C, Warmack R J, Ferrell T L, 1989, \prb{39}
767.
\bibitem{Girard2} Martin O J F, Girard C, Dereux A, 1995, \prl{74} 526.
\bibitem{Girard3} Girard C, Dereux A, 1996, {\it
Rep.\/ Prog.\/ Phys.\/ } {\bf 59} 657.
\bibitem{nonorth} Ward A J, Pendry J B, 1996, \jmo{43} 773.
\bibitem{J+A}Pendry J B, MacKinnon A, 1992, \prl{69} 2772.
\bibitem{JBP} Pendry J B, 1994, \jmo{41} 209.
\bibitem{jackson} Jackson J D, Classical Electrodynamics, 1975, Wiley, 
339-342.
\bibitem{coax} Fischer U C, Zapletal M, 1992, {\it Ultramicroscopy} {\bf 42},
393.
\bibitem{Fischer2} Koglin J, Fischer U C, Brzoska K D, G\"{o}hde W, Fuchs H,
1995, `Photons and Local Probes' 79-92,  Ed. Marti O, M\"{o}ller R, Kluwer
Academic Publishers. 
\end{thebibliography}
\end{document}